\documentclass[]{appolb}
\usepackage{amsmath,amsthm,amssymb,mathrsfs}

\newcommand{\br}[1]{\left(#1\right)}
\newcommand{\ud}[1]{\mathrm{d}#1}
\renewcommand{\exp}[1]{e^{#1}}
\newcommand{\sq}[1]{\left[#1\right]}

\newcommand{\abs}[1]{\left|#1\right|}

\renewcommand{\vec}[1]{\boldsymbol{#1}}
  
  \newcommand{\dd}[1]{\dot{#1}\dot{#1}}

  \newcommand{\pp}{\mathcal{P}}
\newcommand{\qq}{\mathcal{Q}}

\newcommand{\SL}{\mathrm{SL}\br{2,\mathbb{C}}}
\newcommand{\elm}{\mathscr{M}}
\renewcommand{\ell}{\mathscr{L}}
\newcommand{\fl}{{}}
\newcommand{\bss}{\begin{scriptsize}}
\newcommand{\ess}{\end{scriptsize}}
\newcommand{\bsm}{\begin{small}}
\newcommand{\esm}{\end{small}}
\eqsec
\begin{document}

\title{Modified relativistic rotator.\footnote{Presented   at  XLIX Cracow School of Theoretical Physics,
May 31 - June 10, 2009, Zakopane, Poland} \\Toward classical Fundamental Dynamical Systems\\ consisting of a worldline and a single spinor}

\author{\L{}ukasz Bratek \address{The H. Niewodnicza\'{n}ski Institute of Nuclear Physics, Polish Academy of Sciences, \\ Radzikowskego 152, PL-31342 Krak{\'o}w, Poland}}

\maketitle

\begin{abstract}
  The Author shows how to construct a class of Lagrangians for relativistic dynamical systems described by position and a single spinor. One arrives to it by imposing  three requirements $i$) Hamilton action should be reparametrization invariant,
  $ii$)   the number of  dimensional parameters should be minimal,
    $iii$) the spinor phase should be a cyclic variable.

In more detail in this paper  are discussed  the Lagrangians which depend on position and the spinor's null vector only.
An interesting relation of a Hessian determinant and Casimir invariants for such objects leads to the conclusion that no fundamental objects of this kind exist with worldlines uniquely determinable from the Hamilton action and the initial conditions. This unexpected result poses the general question about existence of classical fundamental dynamical systems with well posed Cauchy problem.
\end{abstract}
%
%
\section{Introduction}

My motivation of finding a classical fundamental dynamical system with well posed Cauchy problem arose from a related task of finding an
ideal device with non-quantum clocking mechanism. To deserve the name 'ideal', the building blocks of such a clock should be mathematical ideals, inhabiting nonmaterial Platonic world of mathematical forms, sometimes considered even more realistic than the material world. As pointed out by Prof. Andrzej Staruszkiewicz, such a clock is a way to study some difficult and not well understood
problems in the theory of relativity, such as the clock hypothesis \cite{bib:astar1}.  The hypothesis says that a moving clock
registers  its proper time even when accelerated. This is the hypothetical clock which is carried by particles for registering lengths of their worldlines.

To my mind, however, for over a century since the formulation of
          the special theory of relativity, there has been no successful construction of an  ideal clock which is  deterministic and fundamental at the same time.
                     In the present  work we will follow Staruszkiewicz construction of his fundamental relativistic rotator to find more general dynamical systems which, apart from being fundamental, would be deterministic. By 'deterministic' we mean a dynamical system whose future states are uniquely determined by its initial state. As concerns the word 'fundamental' we adopt the following definition coined by Staruszkiewicz \cite{bib:astar1}:
           \begin{quote}\textit{A relativistic dynamical  system  is  said to
be fundamental if its Casimir invariants are parameters, not constants of
motion}.\end{quote}
For example, putting aside the fact it is non-relativistic, a steadily rotating rigid top is not a fundamental clock, because its intrinsic angular momentum is not independent of the initial angular velocity.

 In order to construct the most elementary  dynamical system with mass and spin,  moving in Minkowski spacetime in accordance with some relativistically invariant laws of motion, two dimensional parameters are needed, for example, mass $\elm$ and length $\ell$. This is required by the existence of two Casimir invariants of the Poincar\'{e} group: $P^{\mu}P_{\mu}$ with the physical dimension of $\elm^2$, and $W^{\mu}W_{\mu}$ with the  physical dimension of $\elm^4\ell^2$.  $W^{\mu}=-\frac{1}{2}\epsilon^{\mu\alpha\beta\gamma}M_{\alpha\beta}P_{\gamma}$
 is the Pauli-Luba\'{n}ski (space-like) spin pseudovector.
 
 By applying the above definition of a fundamental dynamical system, one obtains two independent constraints that must be satisfied by its Lagrangian.
 The unspecified arbitrary parameters $\elm$ and $\ell$ can now be set by relating them directly to the fixed numerical values of the Casimir invariants. With no loss to generality, this can be done by requiring that
\begin{equation} PP\equiv{}\elm^2, \qquad WW\equiv-\frac{1}{4}\elm^4\ell^2.\label{eq:fundcond}\end{equation}
  These two constraints are referred to as \textit{fundamental conditions}.
 The conditions significantly reduce the enormous variety of   relativistically invariant  actions possible for a dynamical system consisting of given mathematical entities.

The simplest clock is described by a spacetime worldline and a single null direction. This is a mathematical abstraction of  Eulerian rotator consisting of two
point masses connected by a rigid and massless rod.  When free, such a system is expected to move periodically in the center of momentum frame. For the sake of visualization, the clocks's dial can be identified in the  frame with  a large circle on the Riemann sphere of null directions which is the image  of the spatial direction of the conserved Pauli-Luba\'nski spin pseudovector. The null direction moves periodically about this circle, counting the number of times the phase has been increased by $2\pi$. The null direction can be thought of as the clock's hand.
Although the motion of the clock's hand can look nonuniform for an external observer (the clock is conformally distorted), the $2\pi$-period will transform via a Lorentz factor (at least when the clock is in free motion).

The fundamental relativistic rotator constructed in \cite{bib:astar1} provides an example of such an ideal clock. Recently, however, it has been shown, that the Cauchy problem  for this clock is ill posed \cite{bib:bratek}. Thus, ideal classical clocks, if exist both as fundamental and deterministic dynamical systems, must be more complex devices. 

This unexpected result poses the question about the existence of fundamental dynamical systems in general.
It seems improbable that fundamental conditions would always imply ill-posedness of the Cauchy problem, it is therefore necessary  to construct a counterexample. 
For that purpose one can consider a class of dynamical systems consisting of a worldline and a single spinor. A spinor can be imagined as a pair consisting of a null vector and a phase associated with rotation in a spatial plane orthogonal to the null vector. The simplest generalization of the fundamental relativistic rotator is thus obtained by finding a fundamental dynamical system in this extended  class. One of possibilities is that such a system could depend on the spinor's null direction and that the spinor's phase would be a cyclic variable. The other possibility, is to completely neglect the spinor's phase and consider a system consisting of a worldline and a single null vector. The third possibility, which presumably will contain a deterministic and fundamental dynamical system, is to assume only that the phase is a cyclic variable. In the present work we shall investigate only the second possibility.

\section{A Tetrad Method for Spinors and Spinor Invariants}

A spinor $\kappa$ is a two component entity over complex numbers that transforms linearly under the action of $\SL$ group. With each spinor one can associate a null vector $$k^{\mu}=\kappa^{+}\sigma^{\mu}\kappa,$$ where $\sigma^{\mu}$ are four Pauli matrices.
 Two spinors with the same null vector differ by a phase. To complete the space of spinors one can associate with $\kappa$ a mate spinor $\tau$  by requiring that $\kappa^0\tau^1-\kappa^1\tau^0=1$. The antisymmetric form associates a volume element with a pair of spinors and is an invariant of $\SL$ group.

 A spinor $\kappa$ can be uniquely determined by specifying  $4$ real numbers: two
 spherical angles $\theta$ and $\phi$ determining the null direction of $k$, the spinor's magnitude $\Psi=+\sqrt{\kappa^{+}\kappa}$, and the spinor's phase $\Phi$. Then $$ \begin{scriptsize}  \kappa=
  \exp{i \Phi/2}\sqrt{\Psi}\sq{\begin{array}{c}\exp{-i\phi/2}\cos\br{\theta/2}\\ \exp{i\phi/2}\sin\br{\theta/2}\end{array}},\qquad \tau=
  \tfrac{\exp{-i \Phi/2}}{\sqrt{\Psi}}\sq{\begin{array}{c}-\exp{-i\phi/2}\sin\br{\theta/2}\\ \exp{i\phi/2}\cos\br{\theta/2}\end{array}}.\end{scriptsize}$$
 Since  $\tilde{\tau}=\tau+\lambda\exp{-i\nu}\kappa$  also solves the unit volume condition for any real $\lambda,\nu$, the mate spinor $\tau$ is determined up to a transformation which, from the viewpoint of spinor $\sigma$, can be considered as a gauge.
 
 \subsection{Spinor Tetrads and Spinor Gauge Transformations}
      Let $m$ and $\tilde{m}$ be null vectors corresponding to $\tau$ and $\tilde{\tau}$, respectively, then $\tilde{m}^{\mu}=m^{\mu}+{2{\lambda}\cos\br{\nu}a^{\mu}+
  2{\lambda}\sin\br{\nu}b^{\mu}}+\lambda^2k^{\mu}$,
  {where} $m^{\mu}=\tau^+\sigma^{\mu}\tau$, $\tilde{m}^{\mu}=\tilde{\tau}^+\sigma^{\mu}\tilde{\tau}$, and \begin{equation}
  \label{eq:ab}\bss
  a^{\mu}=\frac{1}{2}\br{\tau^{+}\sigma^{\mu}\kappa+\kappa^{+}
  \sigma^{\mu}\tau},\qquad
  b^{\mu}=\frac{1}{2i}\br{\tau^{+}\sigma^{\mu}
  \kappa-\kappa^{+}\sigma^{\mu}\tau},\ess  \end{equation}
  are unit spacelike vectors such that $ab=ak=bk=0$.
  Analogously one defines vectors $\tilde{a}$ and $\tilde{b}$ corresponding to the gauged spinor $\tilde{\tau}$, then
  $
  \tilde{a}^{\mu}=a^{\mu}+\lambda\cos\br{\nu}k^{\mu}$ and $
  \tilde{b}^{\mu}=b^{\mu}+\lambda\sin\br{\nu}k^{\mu}$.

  This simple construction shows that every spinor uniquely determines in Minkowski space  a family of spinor tetrads $\br{k,m,a,b}$. The family is invariant with respect to the action of the group of the following gauge transformation
    \begin{equation}\label{eq:gauge}
  \begin{array}{l}\left.\begin{array}{rll}
  k^{\mu}&\to& \tilde{k}^{\mu}=k^{\mu}\\
  a^{\mu}&\to& \tilde{a}^{\mu}=a^{\mu}+\alpha k^{\mu},\quad \alpha\in\mathbb{R}\\
  b^{\mu}&\to& \tilde{b}^{\mu}=b^{\mu}+\beta k^{\mu},\quad \beta\in\mathbb{R}\\
  m^{\mu}&\to& \tilde{m}^{\mu}=m^{\mu}+2\alpha a^{\mu}+2\beta b^{\mu}+\br{\alpha^2+\beta^2}k^{\mu}.\end{array}\right.
  \end{array}\end{equation}
  A composition of two such transformations with parameters $(\alpha_1,\beta_1)$ and $(\alpha_2,\beta_2)$ is again a gauge transformation with parameters $(\alpha_1+\alpha_2,\beta_1+\beta_2)$.
     
   A spinor tetrad $\br{k,m,a,b}$ forms a basis in Minkowski space. Indeed,
  the following Grammian determinant is nonzero
   $$\bss \mathrm{Det}\left[\begin{array}{cccc}kk& km& ka& kb\\
  mk& mm& ma& mb\\
  ak& am& aa& ab\\
  bk& bm& ba& bb \end{array}\right]=-4\ne0\ess,$$
  where
  $aa=bb=-1$, $ab=am=ak=bm=bk=kk=mm=0$ and $km=2$.
 All scalar products in the above determinant are preserved under gauge transformation \eqref{eq:gauge}.
   Any vector $v$ can be expanded in the spinor tetrad as
  $$v=\tfrac{1}{2}(mv)k+\tfrac{1}{2}(kv)m-(av)a-(bv)b,$$  and similarly, $v=\tfrac{1}{2}(\tilde{m}v)\tilde{k}+
  \tfrac{1}{2}(\tilde{k}v)\tilde{m}-(\tilde{a}v)\tilde{a}-(\tilde{b}v)\tilde{b}$.
 A scalar product of vectors $u$ and $v$ reads
  $uv=\frac{1}{2}(kv)(mu)+\frac{1}{2}(ku)(mv)-(au)(av)-(bu)(bv)$.
 \medskip

What has been found above can be rephrased as follows:  a spinor $\kappa$ uniquely defines a null vector $k$ and a family of unit spatial vectors of the form (compare \cite{bib:astarALG})
  \begin{eqnarray*}&a^{\mu}+\alpha k^{\mu}\qquad b^{\mu}+\beta k^{\mu},&\\
   &\abs{\alpha}<\infty,\quad \abs{\beta}<\infty,\qquad ak=0, \quad bk=0,\quad ab=0.&\end{eqnarray*} The triad $\br{k,a,b}$ can be completed by a null vector $m$
   which is uniquely determined by the conditions $km=2$, $am=0$ and $bm=0$, independently of gauge parameters $\alpha$ and $\beta$. By imposing the requirement that scalar products  be preserved when $\alpha$ or $\beta$ are changed, $m$ must transform as in \eqref{eq:gauge}.

   \subsection{Lorentz scalars for a spinor interacting with its worldline}

    A class of spinor tetrads $(k,m,a,b)$ related by  gauge transformation \eqref{eq:gauge} can be considered as a spacetime realization of a single spinor. This viewpoint makes straightforward the task of finding all functionally independent Lorentz scalars for a dynamical system consisting of  a single spinor interacting with its worldline.

    All Lorentz scalars formed from a spinor  and its first derivatives should be invariants of gauge transformation \eqref{eq:gauge}. It is assumed that the spinor couples minimally with its worldline. It means that the Lagrangian should be a function of gauge invariants formed from scalar products of  $\dot{x}$ (the tangent to the spinor's worldline), spinor tetrad $k,m,a,b$, and velocities $\dot{k},\dot{m},\dot{a},\dot{b}$.

There is $16$ nonzero scalar products formed from first derivatives of a spinor tetrad: $a\dot{b}$, $ a\dot{k}$, $ a\dot{m}$, $ b\dot{k}$, $ b\dot{m}$, $ \dot{a}\dot{a}$, $
\dot{a}\dot{b}$, $ \dot{a}\dot{k}$, $
\dot{a}\dot{m}$, $ \dot{b}\dot{b}$, $ \dot{b}\dot{k}$, $ \dot{b}\dot{m}$, $
\dot{k}\dot{k}$, $ \dot{k}\dot{m}$, $ \dot{m}\dot{m}$, $ m\dot{k}$.
On decomposing velocities $\dot{k}$, $\dot{m}$, $\dot{a}$, $\dot{b}$ in the basis $\br{k,m,a,b}$, one infers that the scalars quadratic in velocities can be expressed as binomials of  scalars linear in these velocities, e.g., $\dot{k}\dot{k}=-((a\dot{k})^2+(b\dot{k})^2)$.
This reduces the number of scalars  to $6$:
$a\dot{b}$, $a\dot{k}$, $a\dot{m}$, $b\dot{k}$, $b\dot{m}$, $m\dot{k}$.
However, they are not functionally independent. For example, in a gauge in which $k=K\sq{1,\vec{n}}$, $m=K^{-1}\sq{1,-\vec{n}}$, $a=\sq{0,\vec{a}}$, $b=\sq{0,\vec{a}\times\vec{n}}$, $\vec{n}\vec{n}=\vec{a}\vec{a}=1$, $\vec{a}\vec{n}=0$, there is $a\dot{m}\cdot{}b\dot{k}=a\dot{k}\cdot{}b\dot{m}$ (not a gauge invariant).

The number of basic Lorentz scalars  extends by taking into account the interaction of spinor $\kappa$ with its worldline. The tangent to the worldline, $\dot{x}$, decomposes in the spinor basis as
$\dot{x}=\tfrac{1}{2}(m\dot{x})k+\tfrac{1}{2}(k\dot{x})m-(a\dot{x})a-(b\dot{x})b$. This gives $4$ interaction scalars $m\dot{x}$, $k\dot{x}$, $a\dot{x}$, $m\dot{x}$.
It is clear that other possible interactions $\dot{k}\dot{x}$, $\dot{m}\dot{x}$, $\dot{a}\dot{x}$, $\dot{b}\dot{x}$ or even $\dot{x}\dot{x}$ add nothing new, since they can be expressed as functions of the already found scalars, e.g., $\dot{x}\dot{x}=(k\dot{x})(m\dot{x})-(a\dot{x})^2-(b\dot{x})^2$ or $\dot{k}\dot{x}=\frac{1}{2}(k\dot{x})(m\dot{k})-(a\dot{k})(a\dot{x})-(b\dot{k})(b\dot{x})$.

Accordingly, there is $10$ basic Lorentz scalars describing a spinor interacting with its worldline.
These scalars
transform under  gauge transformation \eqref{eq:gauge} as shown in the following table
$$
\begin{array}{|c|}
\hline
\begin{array}{@{}l|r@{}}
\begin{array}{c}
a\dot{k}\to a\dot{k}\\
b\dot{k}\to b\dot{k}\\
k\dot{x}\to k\dot{x}
\end{array}&
\begin{array}{c}
a\dot{x}\to a\dot{x}+\alpha k\dot{x}\\
b\dot{x}\to b\dot{x}+\beta k\dot{x}\\
a\dot{b}\to a\dot{b}- \alpha b\dot{k}+\beta a\dot{k}\\
m\dot{k}\to m\dot{k}+2\alpha{}a\dot{k}+2\beta{}b\dot{k}\\
m\dot{x}\to m\dot{x}+2\alpha a\dot{x}+2\beta b\dot{x}+(\alpha^2+\beta^2)k\dot{x}\\
\end{array}\end{array}\\
\hline
a\dot{m}\to a\dot{m}-\br{\alpha^2-\beta^2}a\dot{k}-\alpha m\dot{k}+2\beta a\dot{b}-2\alpha\beta b\dot{k}-2\dot{\alpha}\\
b\dot{m}\to b\dot{m}+\br{\alpha^2-\beta^2}b\dot{k}
-2\alpha a\dot{b}-\beta m\dot{k}-2\alpha\beta a\dot{k}-2\dot{\beta}.\\
\hline
\end{array}$$
Spinor invariants are Lorentz scalars which are independent of  gauge parameters $\alpha$ and $\beta$.
Scalars $a\dot{m}$ and $b\dot{m}$ must be rejected since they contain $\dot{\alpha}$ and $\dot{\beta}$ which cannot be removed using the remaining scalars. As concerns the $5$ scalars in the upper right block (denoted respectively by $J_1$, $\dots$ $J_5$), the most general binomial in $\alpha$ and $\beta$ is
$$\sum\limits_{i=1}^{4}
\sq{c_i+\sum\limits_{j=i}^{4}d_{ij}J_j}J_i+c_5J_5.$$ By equating to zero the coefficients  standing at $\alpha$, $\beta$, $\alpha\beta$, $\alpha^2$, and $\beta^2$ in the binomial, one obtains a system of $5$ linear equations of the form $AV=0$ for a $15$-dimensional vector $V=\{c_1,\dots,c_5,d_{11},\dots,d_{14},d_{22},d_{23},\dots,d_{44}\}$. Since the rank of matrix $A$ is $5$, there is $10$ nontrivial null spaces.  Two of the 10 corresponding spinor invariants are identically zero. The rank of a rectangular matrix consisting of first derivatives of the remaining $8$ spinor invariants with respect to the $10$ basic scalars displayed in the above table is $3$, therefore only $3$ invariants are functionally independent. Together with the invariants in the upper-left block in this table, they give $6$ functionally independent spinor invariants.
These findings can be summarized as follows
\begin{quote}\item
\textit{There exist $6$ functionally independent Lorentz scalars for a system described by a spinor represented by a spinor tetrad $k,m,a,b$ interacting with its worldline. These invariants read}
$$\begin{array}{l}
\iota_1=a\dot{k},\qquad
\iota_2=b\dot{k},\qquad
\iota_3=k\dot{x}\\
\iota_4=(k\dot{x})(m\dot{x})-(a\dot{x})^2-(b\dot{x})^2\equiv \dot{x}\dot{x}\\
\iota_5=\frac{1}{2}(k\dot{x})(m\dot{k})-(a\dot{k})(a\dot{x})-(b\dot{k})(b\dot{x})\equiv \dot{k}\dot{x}\\
\iota_6=(a\dot{x})(b\dot{k})-(a\dot{k})(b\dot{x})+(a\dot{b})(k\dot{x}).
\end{array}$$
\end{quote}
Equation \eqref{eq:ab} implies that when the spinor phase changes by an angle $\Delta_{\Phi}$, vectors $a$ and $b$ get rotated through the same angle
$$ a^{\mu}\to a^{\mu}\cos{\Delta_{\Phi}} -b^{\mu}\sin{\Delta_{\Phi}} ,\quad
b^{\mu}\to a^{\mu}\sin{\Delta_{\Phi}} +b^{\mu}\cos{\Delta_{\Phi}}.$$
The spinor invariants change correspondingly as
$$\begin{array}{ll}
\iota_1\to  \iota_1\cos{\Delta_\Phi} -\iota_2\sin{\Delta_\Phi}&\iota_3\to \iota_3\\
\iota_2\to \iota_1\sin{\Delta_\Phi}+\iota_2\cos{\Delta_\Phi}&\iota_4\to \iota_4\\
\iota_6\to \iota_6+\iota_3\dot{\Delta}_{\Phi}&\iota_5\to \iota_5.
\end{array}$$

 \section{Construction of Lagrangians}

 As follows from the previous section, there is only $5$ functionally independent spinor invariants which are explicitly phase-independent: $\iota_1^2+\iota_2^2$, $\iota_3$, $\iota_4$, $\iota_5$ and $\iota_6$  (the latter depends on the first derivative of the phase).
However, not every combination of the invariants is suitable for a relativistically invariant Hamilton's action. Such an action must be reparametrization invariant.
 This reduces the number of useful phase independent spinor invariants to $4$:
$$
\begin{array}{l|l}
I_1=\frac{(a\dot{k})^2+(b\dot{k})^2}{(k\dot{x})^2}\equiv -\frac{\dot{k}\dot{k}}{{(k\dot{x})^2}} & I_3=\frac{(k\dot{x})(m\dot{k})-2(a\dot{k})(a\dot{x})-2(b\dot{k})(b\dot{x})}{
{2k\dot{x}\sqrt{\dot{x}\dot{x}}}}\equiv \frac{\dot{k}\dot{x}}{k\dot{x}\sqrt{\dot{x}\dot{x}}}
\\
I_2=\ \frac{(a\dot{x})(b\dot{k})-(a\dot{k})(b\dot{x})+(a\dot{b})(k\dot{x})
}{k\dot{x}\sqrt{\dot{x}\dot{x}}} &I_4=\frac{k\dot{x}}{\sqrt{\dot{x}\dot{x}}}
\end{array}
$$
 Together with the worldline scalar $I_0= \dot{x}\dot{x}$ they are suitable to construct the most general action for a dynamical system consisting of a single spinor interacting with its worldline. Spinor phase independence of the Lagrangian ensures the existence of an additional integral of motion, independently of relativistic symmetries. This explains why spinor invariants explicitly depending on the spinor's phase has been rejected.

  As we have established earlier, an action of a relativistic dynamical system should possess at least two dimensional parameters: mass $\elm$ and length $\ell$. If no other dimensional parameters are assumed, the spinor invariant $I_4$, which would introduce its own physical dimension, must be rejected. This way one is led to the following class of actions
 \begin{equation}\label{eq:spinorclass}\mathcal{S}\sq{x,\kappa}=
 -\elm\int\ud{s}\sqrt{I_0}\,\mathcal{F}(\ell^2I_1,\ell{}I_2,\ell{}I_3).\end{equation}
 This class of Lagrangians, labeled by function $\mathcal{F}$, has been arrived to by imposing only three transparent requirements on a Hamilton action of a dynamical system consisting of a spinor interacting with its worldline. Namely, $i$) the action should be reparametrization invariant, $ii$) the number of  dimensional parameters should be minimal (that is, two), and $iii$) the spinor phase should be a cyclic variable.

 \section{A single null vector interacting with its worldline}

 From now on we shall be discussing a subclass of actions \eqref{eq:spinorclass} depending on
 the null vector of spinor $\kappa$  and its worldline
 \begin{equation}
\fl -\elm\int\sqrt{\dot{x}\dot{x}}\,F\br{{\pp},{\qq}}\ud{\tau},
\qquad {\pp}=\ell\frac{\dot{k}\dot{x}}{k\dot{x}\sqrt{\dot{x}\dot{x}}}, \quad  {\qq}=-\ell^2\frac{\dot{k}\dot{k}}{\br{k\dot{x}}^2}.\label{eq:action}\end{equation}
This class of dynamical systems was studied in more detail in \cite{bib:bratek2} as extension of the class of rotators studied in \cite{bib:bratek} that includes the fundamental relativistic rotator \cite{bib:astar1}. Here we recall only the most important results.

The invariance with respect to space-time translations and space-time rotations, implies conservation of momentum $P^{\mu}$ and angular momentum $M_{\mu\nu}$, respectively. The Casimir invariants of the Poincar\'{e} group are
\begin{eqnarray*}\fl \phantom{WWW}
     PP&=&\elm^2\,\left( \left( F - {}{\pp\,F_{,\pp}}\,  \right) \,
     \left( F - {}{\pp\,F_{,\pp}}\,  -
       4\,{}{\qq \,F_{,\qq}}\, \right)  -
    {{\qq\,}{F_{,\pp}}}^2 \right),\\
    \fl \phantom{WWW}
           WW&=&
-\elm^4\,{\ell}^2 \qq\,{\left( {{}{F_{,\pp}}}^2 +
      2\,{}{F_{,\qq}}\,\left( F - {\pp\,}{F_{,\pp}}\,
         \right)  \right) }^2.\end{eqnarray*}
By applying fundamental conditions \eqref{eq:fundcond} one obtains two independent differential equations for unknown function $F$. There is no apparent reason for these two unrelated differential equations to  have a common solution. Remarkably enough, two such solutions are possible  giving rise to two fundamental dynamical systems. These solutions can be found by means of Legendre transformations and they read \cite{bib:bratek2}
\bsm \begin{eqnarray*}
F(\pp,\qq)=\pm\sqrt{
\br{1\pm\sqrt{\qq}}
\br{1+\frac{\pp^2}{\qq}}
},\qquad
F(\pp,\qq)={\nu \pp\pm
\sqrt{{1\pm\sqrt{\qq}}-\nu^2\qq}}, \end{eqnarray*}\esm  where $\nu\in\mathbb{R}$ is a dimensionless  integration constant of fundamental conditions.

 Before presenting the  actions corresponding to these solutions, it  will be instructive to discuss some implications of the following relationship between a Hessian determinant associated with action \eqref{eq:action} and a Jacobian determinant of an $F$-dependent mapping $\br{PP\br{\pp,\qq},WW\br{\pp,\qq}}$ derived in   \cite{bib:bratek2}
\begin{equation}\label{eq:HessianJacobian}\det{\mathcal{H}}=\mathcal{K}\cdot
\frac{F-\pp\,F_{,\pp}}{F_{,\pp}\br{\pp^2+\qq}-\pp{}F}
\cdot\,\left|\frac{\partial\br{PP,WW}}{\partial\br{\pp,\qq}}\right|.
\end{equation}
Here, $\mathcal{H}$ is a matrix of second derivatives of the Lagrange density in action \eqref{eq:action} with respect to generalized velocities associated with all dynamical degrees of freedom. Kinematical factor $\mathcal{K}$ is a function of  generalized velocities and it is independent of function $F$.

In the distinguished case when $F_{,\pp}\br{\pp^2+\qq}-\pp{}F=0$, that is, when $F=\sqrt{1+\frac{\pp^2}{\qq}}S\br{\qq}$ with $S$ being some function,
the Jacobian determinant vanishes, but not necessarily does the Hessian determinant (indeterminate form $\tfrac{0}{0}$).
In this case the Casimir invariants are functionally dependent: $PP=\elm^2S\br{S-4\qq{}S'}$ and $WW=-\br{2\elm^2\ell{}S\sqrt{\qq}S'}^2$,  while $\det{\mathcal{H}}\propto\frac{\qq{}S^3S'}{\br{\pp^2+\qq}^2}
\br{2\qq\br{S'}^2+S\br{S'+2\qq{}S''}}$, that is, $\det{\mathcal{H}}\propto{}S^3S'\br{PP}'$ or
$\det{\mathcal{H}}\propto{}S^2\br{WW}'$. Then $\det{\mathcal{H}}\ne0$ unless fundamental conditions are imposed.
In all other cases, when $F_{,\pp}\br{\pp^2+\qq}-\pp{}F\neq 0$,  vanishing of the Hessian determinant is equivalent to vanishing of the Jacobian determinant (if $F-\pp{}F_{,\pp}=0$ then $WW=\elm^2{\ell^2}PP$, which is unphysical).

The above observations lead to the central conclusion that a dynamical system defined by action \eqref{eq:action} is defective
when it is fundamental, since then $\det{\mathcal{H}}=0$ (the implications of condition $\det{\mathcal{H}}=0$ will become clear later).

\medskip
Hamilton actions of the fundamental systems corresponding to the previously found solutions read, respectively,
\bsm\begin{equation}
\label{eq:StarlikeAction}\fl S=-\elm\int\ud{\tau}\sqrt{\dot{x}\dot{x}}\sqrt{
 \sq{1-\frac{(\dot{k}\dot{x})(\dot{k}\dot{x})}
 {(\dot{x}\dot{x})(\dot{k}\dot{k})}}
 \sq{1\pm\sqrt{-\ell^2\frac{\dot{k}\dot{k}}{\br{k\dot{x}}^2}}}}
 \end{equation}
 and
 \begin{equation}
\label{eq:myaction}
\fl
S_{\nu}=-\elm\int\ud{\tau}\sqrt{\dot{x}\dot{x}}
\br{
 \sqrt{1\pm\sqrt{-\ell^2\frac{\dot{k}\dot{k}}{\br{k\dot{x}}^2}}+
 \nu^2\,{\ell^2}\frac{\dot{k}\dot{k}}{\br{k\dot{x}}^2}  }+\nu\,{\ell}\frac{\dot{k}\dot{x}}{k\dot{x}\sqrt{\dot{x}\dot{x}}} }.\end{equation}\esm
  In contrast to dynamical system  defined by \eqref{eq:StarlikeAction}, which has $6$ dynamical degrees of freedom,
    the system defined by action \eqref{eq:myaction} must be treated as having only $5$ dynamical degrees of freedom, since the magnitude of null vector $k$ in this case is a gauge variable.
 Indeed, for any function $\psi(\tau)$ $$S_{\nu}[x,\exp{\psi}k]=S_{\nu}[x,k]-\elm\ell\,\nu\,\psi(\tau).$$  Since the Lagrangians corresponding to actions $S_{\nu}[x,\exp{\psi}k]$ and $S_{\nu}[x,k]$ differ by a total derivative, the form of equations of motion is left unchanged.  This means that the amplitude of null vector $k^{\mu}$ separates completely from the dynamics of other degrees of freedom and does not influence them at all, therefore it can be completely ignored.  As a result, the dynamical system defined by action \eqref{eq:myaction} depends on position and a null direction only, similarly as the fundamental relativistic rotator.
   The rank of Hessian matrix for action \eqref{eq:myaction} equals $4$ (which is less than the number of dynamical degrees of freedom even when the amplitude of $k^{\mu}$ is not taken into account).

\medskip
 What the defectiveness of the found fundamental systems  means in practice? Firstly, it should be recalled that the necessary condition for the existence of Hamiltonian mechanics for a dynamical system
described by a general Lagrangian $L(v,q)$ depending on generalized coordinates $q$ and velocities $v$ compatible with constraints, is that for fixed $q$ the set of equations $p(v,q)=\frac{\partial{}L}{\partial{}v}(q,v)$ defining momenta $p$, should be a diffeomorphism of spaces of momenta $p$ and of velocities $v$. In particular, this set of equations should be uniquely solvable  for velocities, $v=v(q,p)$. This is possible, provided that the following Hessian determinant is nonzero $$\det\sq{\frac{\partial{}^2L}{\partial{}\dot{q}^i\partial\dot{q}^j}}\ne0,$$ otherwise the Legendre transform leading from the Lagrangian to the Hamiltonian would  not be well defined.
Secondly, the above condition  can be equivalently viewed as  necessary for  unique dependence of accelerations on the initial data. The
Euler-Lagrange equations for $L$ can be recast in the general form
$$\frac{\partial{}^2L}{\partial{}\dot{q}^i\partial\dot{q}^j}\ddot{q}^j=Z(q,\dot{q},t),$$ with some function $Z$. Therefore, the vanishing of the Hessian determinant would not only mean that accelerations could not be algebraically determined from the positions $q$ and their derivatives, but also that equations of motion could not be reduced to the canonical form  $\dot{y}=Y(y,t)$, where $y=(q,\dot{q})$, for which the general textbook results about existence and uniqueness are derived for solutions of ordinary differential equations.

\subsection{Singular motion of  Fundamental Relativistic Rotator.}
 It is best to illustrate the consequences of vanishing of Hessian determinat with the behavior of the fundamental relativistic rotator.
 
 The action of fundamental relativistic rotator  \cite{bib:astar1} is obtained from  \eqref{eq:myaction} by taking the formal limit $\nu\to0$
 \begin{equation}\label{eq:AstarAction} S_{\nu=0}=-\elm\int\ud{\tau}\sqrt{\dot{x}\dot{x}}
\sqrt{1+\sqrt{-\ell^2\frac{\dot{k}\dot{k}}{\br{k\dot{x}}^2}}}.\end{equation}
The same action was obtained earlier in quite a different context in \cite{bib:segal} as a geometrical model  of a spinning massive particle. 

Free motion of the rotator has been found in a fully covariant form in \cite{bib:bratek}. Its parametric description reads
\begin{equation}\label{eq:gensol}\fl
x^{\mu}(t)=\frac{P^{\mu}}{\elm}t+\frac{\ell}{2}r^{\mu}(t)+x^{\mu}(0),\quad  k^{\mu}\br{t}=\frac{P^{\mu}}{\elm}+\frac{\dot{r}^{\mu}(t)}{{\sqrt{-\dot{r}(t)\dot{r}(t)}}},
\end{equation}
where
$$r^{\mu}(t)=N^{\mu}\sin{\phi(t)}+
\frac{\epsilon^{\mu\nu\alpha\beta}N_{\nu}W_{\alpha}P_{\beta}}{
\frac{1}{2}\elm^3\ell}\cos{\phi(t)}.$$
Constant vectors $P^{\mu}$, $W^{\mu}$ and $N^{\mu}$ satisfy the following conditions
$PP=\elm^2$, $WW=-\frac{1}{4}\elm^4\ell^2$, $WP=0$, $NN=-1$, $NW=0$, and $NP=0$. $P^{\mu}$ is the (conserved) momentum of the center of momentum frame, $t$ is the proper time in this frame, and $W^{\mu}$ is the (conserved) spin pseudovector.

Function $\phi(t)$ describes the angular position of the "pointer" $k^{\mu}(t)$ in the center of momentum frame. The angular velocity with which $k^{\mu}$ moves on the unit sphere of null directions in this frame is
$$\abs{\frac{\ud{\phi}}{\ud{t}}}=\frac{2}{\ell}\tanh{\Psi}, \qquad \exp{2\Psi}\equiv{\sqrt{-
\ell^2\frac{\dd{k}}{\br{\dot{x}k}^2}}+1}.$$ This function is such that
$0<|{\dot{\phi}(t)}|<\frac{\ell}{2}$ and otherwise arbitrary. For a well behaving dynamical system, ${{\phi}(t)}$ would be a linear function of parameter $t$. From the physical standpoint this arbitrariness of $\phi(t)$  is unacceptable, since it would mean that a dynamical system could accelerate or decelerate at will without apparent cause. Putting this differently, function $\phi(t)$ is not determined uniquely from equations of motion and initial conditions. In this sense, fundamental relativistic rotator is not a deterministic dynamical system. 
One  expects that motion of fundamental dynamical systems defined by actions \eqref{eq:StarlikeAction} and \eqref{eq:myaction} is similarly defective.

This indeterministic behavior originates neither from reparametrization invariance of action \eqref{eq:AstarAction} nor its invariance with respect to rescaling of the null vector $k^{\mu}$ by arbitrary function.
It is inherent in the particular form of the
Hamilton's action of the fundamental relativistic rotator. For example, a Hessian determinant calculated for $5$ dynamical degrees of freedom (the amplitude of $k^{\mu}$ is not taken into account) of the dynamical system defined by action
 $$S=-\elm\int\ud{\tau}\sqrt{\dot{x}\dot{x}}\,f(\qq),\qquad \qq=-\ell^2\frac{\dot{k}\dot{k}}{\br{k\dot{x}}^2},$$
 reads \cite{bib:bratek}
  $$\det{\mathcal{H}_{5d.o.f}}=\mathcal{K}{f(Q)^3f'(Q)^2}\br{1+2Q\br{
\frac{f'(Q)}{f(Q)}+\frac{f''(Q)}{f'(Q)}}},\qquad \mathcal{K}_{,f}\equiv0.$$
It is nonzero for arbitrary nonconstant function $f(\qq)$ different from function $c_1\sqrt{1+c_2\sqrt{\qq}}$ ($c_1=1$ and $c_2=\pm1$ to have $PP=\elm^2$ and $WW=-\tfrac{1}{4}\elm^4\ell^2$). For such 
  $f$'s one obtains deterministic systems (in the sense that $\det{\mathcal{H}}\ne0$), however, not fundamental -- their mass and spin are functions of initial conditions.

\section{Summary}

The requirement that Casimir invariants of the Poincar\'{e} group should be parameters rather than constants of motion, implies indeterministic behavior of classical dynamical systems described by position and a single null vector. In particular, this concerns the fundamental relativistic rotator. 

 Ideal classical clocks, if exist, must be more complicated devices. For more complicated systems, it seems  improbable that satisfaction of fundamental conditions  would always imply ill-posedness of the Cauchy problem. It is therefore necessary to find a counterexample.

For that purpose, one can consider dynamical systems consisting of a single spinor interacting with its worldline. In this work it has been shown how to construct the Lagrangian for such a system. The system is described by three Lorentz scalars, thus more than the number of Casimir invariants of the Poincar'e group. One therefore expects to have no similar correspondence between vanishing of Hessian determinant and functional dependence of Casimir invariants as that observed for systems described by action \eqref{eq:action} with two Lorentz invariants. It would be also interesting to show in general whether or not fundamental dynamical systems with two Lorentz scalars have always vanishing Hessian determinant. 

The existence of deterministic fundamental dynamical systems consisting of a worldline and a single spinor is still an open question. However, I have already found an indication that action
{}
\begin{scriptsize}
$$\mathcal{S}\sq{x,\kappa}=
 -\elm\int\ud{s}\sqrt{\dot{x}\dot{x}}\,\mathcal{F}
 \br{-\ell^2\frac{\dot{k}\dot{k}}{{(k\dot{x})^2}},\ell{}\frac{(a\dot{x})(b\dot{k})-(a\dot{k})(b\dot{x})+(a\dot{b})(k\dot{x})
}{k\dot{x}\sqrt{\dot{x}\dot{x}}} }$$\end{scriptsize}\noindent
leads to at least one family of fundamental systems with vanishing Hessian determinant. It seems that to find a deterministic and fundamental system with spinor one needs to consider full Hamilton action \eqref{eq:spinorclass} which is a very challenging endeavor.  

\begin{small}

\end{small}
\end{document}